\documentclass{article}

\usepackage{arxiv}

\usepackage[utf8]{inputenc} 
\usepackage[T1]{fontenc}    
\usepackage{hyperref}       
\usepackage{url}            
\usepackage{booktabs}       
\usepackage{amsfonts}       
\usepackage{nicefrac}       
\usepackage{microtype}      
\usepackage{lipsum}		
\usepackage{graphicx}
\usepackage[numbers]{natbib}
\usepackage{doi}
\usepackage{subcaption}
\usepackage{multirow}
\usepackage{bbding}
\usepackage{colortbl}
\usepackage{xcolor}
\usepackage{array}
\definecolor{darkgreen}{RGB}{0, 128, 0}
\newcommand{\tick}{\textcolor{darkgreen}{\Checkmark}}
\newcommand{\cross}{\textcolor{red}{\XSolidBrush}}
\newcommand{\asterisk}{\textcolor{orange}{\FourStar}}

\title{Methods to assess the UK government's current role as a data provider for AI}

\date{} 					

\author{ \href{https://orcid.org/0009-0008-3969-2514}{\includegraphics[scale=0.06]{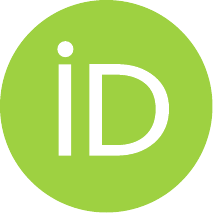}\hspace{1mm}Neil Majithia}\thanks{Alternative email address: \texttt{neil.majithia@live.co.uk}} \\
	Open Data Institute (ODI)\\
	London\\
	\texttt{neil.majithia@theodi.org} \\
	\And
	\href{https://orcid.org/0000-0003-1722-947X}{\includegraphics[scale=0.06]{orcid.pdf}\hspace{1mm}Elena Simperl} \\
	Open Data Institute (ODI)\\
	London\\
	\texttt{elena.simperl@theodi.org} \\
}

\renewcommand{\shorttitle}{Methods to assess the UK government's current role as a data provider for AI - ODI}

\hypersetup{
pdftitle={Methods to assess the UK government's current role as a data provider for AI},
pdfauthor={Neil Majithia},
pdfkeywords={Generative AI, Data-centric AI, AI public policy, Open Government Data, Training data},
}

\begin{document}
\maketitle

\begin{abstract}
    Governments typically collect and steward a vast amount of high-quality data on their citizens and institutions, and the UK government is exploring how it can better publish and provision this data to the benefit of the AI landscape. However, the compositions of generative AI training corpora remain closely guarded secrets, making the planning of data sharing initiatives difficult. To address this, we devise two methods to assess UK government data usage for the training of Large Language Models (LLMs) and ‘peek behind the curtain’ in order to observe the UK government’s current contributions as a data provider for AI.
    
    The first method, an ablation study that utilises LLM ‘unlearning’, seeks to examine the importance of the information held on UK government websites for LLMs and their performance in citizen query tasks. The second method, an information leakage study, seeks to ascertain whether LLMs are aware of the information held in the datasets published on the UK government’s open data portal data.gov.uk. Our findings indicate that government websites are important data sources for AI (though importance varies across subjects) while data.gov.uk is not. 
    
    This paper serves as a technical report, explaining in-depth the designs, mechanics, and limitations of the methods. It is accompanied by a report on the ODI website\footnote{The accompanying ODI report, "The UK government as a data provider for AI" is available at \url{https://theodi.org/insights/reports/the-uk-government-as-a-data-provider-for-ai}} in which we summarise the experiments and key findings, interpret them, and build a set of actionable recommendations for the UK government to take forward as it seeks to design AI policy. While we focus on UK open government data, we believe that the methods introduced in this paper present a reproducible approach to tackle the opaqueness of AI training corpora and provide organisations a framework to evaluate and maximize their contributions to AI development.\footnote{We release the code and data used in this paper at \url{github.com/NevadaM/DCAI\textunderscore ODI\textunderscore GovAsADataProvider}}
    
    \scriptsize{Copyright \copyright  2024, Open Data Institute (\url{www.theodi.org}). All rights reserved.}
\end{abstract}


\section{Introduction}
Artificial Intelligence (AI) models continue to dominate the technological landscape, but despite their ubiquity across contexts, many questions remain regarding their underlying training corpora whose contents are often closely held secrets \cite{Hardinges2024We}.

This has implications in the world of public policy. Governments are likely to be extremely important actors in the emergent AI future, with the ODI's 2024 white paper ‘Building a better future with data and AI’ \citep{theodiBuildingBetter} presenting the many facets by which the UK government should plan its  future role in the development and governance of AI. In the white paper, we propose that the UK government can put into practice its "pro-innovation" stance by playing the role of a \textbf{data provider for AI}. In doing so, the government has the opportunity to be a "catalyst for AI development" by ensuring broad access to high-quality, structured data \citep{GPAIroleofgov}.  

The government is already taking steps towards being a data provider for AI by developing and testing proof of concepts like the National Data Library \citep{theodiNDLinput}. Before these initiatives are implemented, it is important to understand the current state of the government in this aspect: while a lot of government data sources are open and freely available to developers right now, are they actually used for the training of AI models, and if so, to what extent? In other words:

\begin{quote}
    \textbf{Research Question}: \emph{To what extent do UK government data sources contribute to the performance of AI models?}
\end{quote}

To answer the question, we introduce, implement, and perform initial evaluations of two methods for this purpose: an ablation study (section \ref{sec2}), and an information leakage study (section \ref{sec3}). Each experiment focuses on Large Language Models (LLMs) only.

Although the research question is specific to the UK government, the methods introduced here have broader applicability, offering a framework for examining the role of other organisations and datasets in AI training corpora. Therefore, we hope this work not only provides insights into the UK government’s current role but also showcases methods that can be meaningfully contributive to the field of data-centric AI research. 

\subsection{UK government data and its use for AI}
The UK government holds a multitude of data that could be useful for AI models. Beyond \href{https://www.ons.gov.uk/}{official statistics releases} and \href{https://www.nationalarchives.gov.uk/}{national archives}, in this paper we examine two other UK government data sources: government websites and government open data published on data.gov.uk.

\begin{enumerate}
    \item \textbf{Government websites} \footnote{Such as \url{https://www.gov.uk/universal-credit/eligibility}} contain textual information on UK government policies, services, and guidance, written in plain, accessible English. As a structured collection of authoritative knowledge, these websites can enhance the performance of LLMs by offering accurate and trustworthy information about the governance and overall lifestyle of the UK. 
    \item \textbf{data.gov.uk} is the government's central open data platform, hosting a wide variety of datasets,  ranging from public health and policing statistics to environmental and economic indicators. Unlike websites, these datasets are often presented in numerical or tabular formats, providing quantitative insights into the UK government’s activities and its citizens' lives.
\end{enumerate}

These data sources can play a critical role in supporting LLMs when they are employed in public service tasks. Such tasks involve a citizen of the UK asking an LLM about government policies or services, often in the context of their personal circumstances. Historically, these queries would have been answered through interaction with civil service help desks or static searches of government websites. LLMs, however, offer the potential for multi-round dialogues that enhance accessibility and responsiveness \citep{complandmultiround}.

\section{The importance of government websites for LLMs - an ablation study}
\label{sec2}
Government websites provide a wealth of textual information about UK policies, welfare programmes, and public services. Written in accessible English, this information is structured and reliable, making it an ideal resource for training large language models (LLMs). These websites are particularly valuable for answering citizen queries as they offer authoritative guidance on nuanced topics such as eligibility criteria for benefits or the interaction between various welfare schemes.

We adopt a counterfactual approach: what would happen if LLMs didn't have access to UK government websites? Specifically, we ask:
\begin{quote}
    \emph{How much would the performance of LLMs in citizen query tasks suffer if UK government websites were not in their training corpora? }
\end{quote}

\subsection{Candidate Methods in Related Work}
\subsubsection{Influence functions}
Influence functions are statistical techniques that can measure the contribution of chosen samples of training data towards the predictions of a model. In our case, they could be used to numerically estimate how valuable government websites are to LLMs’ responses to citizen queries and therefore determine the extent to which models would be worse off if said websites were not in their training corpora.

Both \citet{koh2020understandingblackboxpredictionsinfluence} and  \citet{grosse2023studyinglargelanguagemodel} explore the use of influence functions in machine learning, providing mathematical formulations, python implementations, and preliminary results. However, both works admit the heavy computational load required to perform influence function calculation on account of the need to compute or at least estimate the inverse Hessians of the targeted models, something difficult to even conceptualise when models reach the size and dimensionality of current-day generative AI. Recently however, \citet{choe2024dataworthgptllmscale} build and test a an efficient influence function measurement framework built upon gradient projection strategies; therefore, in the foreseeable future, influence function methods may be more feasible for generative AI work. 

\subsubsection{Ablation studies}

As a counterfactual, the research question can be developed beyond a simple thought exercise via an ablation study, which could involve removing government website data from the training corpora of LLMs, retraining them on the new corpora, and assessing how the removal affects the models’ function and performance in citizen query tasks. Intuitively, this sounds computationally intensive and requires easy access to the full training corpora of LLMs (which often remain proprietary \citep{Hardinges2024We}), as well as a replication of the original LLM training protocol to be performed again with the new corpora. 

Instead, ‘unlearning’ methods built for the removal of harmful or copyrighted output from AI models (introduced in \citet{yao2024largelanguagemodelunlearning}) don’t require retraining, instead performing a targeted reverse-fine-tuning on the models so that they ‘forget’ certain parts of their training corpora.

Yao et al.’s technique, especially when training LLMs to forget copyrighted material, provides a robust method that seems surgically accurate and minimally disruptive to models’ overall performance. Their method has been adopted for an entirely different purpose in \citet{lu2024eraserjailbreakingdefenselarge}, while other unlearning methods have also been designed across literature \citep{hu2024exactefficientunlearninglarge}.  

\subsection{Methods}
\subsubsection{Ablation method}
Yao et al.'s unlearning method is analagous to a sort of fine-tuning of a subject LLM $\theta$ on some 'target' dataset $X^{target}$, but in the reverse direction to typical fine-tuning so that the model 'forgets' the data rather than learn it. In other words, it is an implementation of gradient descent, where for each step of the process $t$, $\theta$ is gradually transformed so that loss on the target dataset increases:
\begin{equation}
    L(\theta_{t+1}(X^{target})) > L(\theta_{t}(X^{target}))
\end{equation}

where $L(\dots)$ is the cross-entropy loss function. However an important, secondary objective of Yao et al.'s unlearning method is that despite the gradient ascent on target data, the subject LLM should preserve language capabilities and its knowledge of non-target data $X^{safe}$. So, each step of the process also aims for loss on the safe dataset to remain approximately the same:
\begin{equation}
    L(\theta_{t+1}(X^{safe})) \approx L(\theta_{t}(X^{safe}))
    \label{eqn2}
\end{equation}

To achieve both of these objectives, Yao et al.'s unlearning process is designed as a training procedure composed of three loss functions. These are best explained in the original paper, but essentially work to perform gradient ascent on target data while also mitigating negative effects on safe data (rearranging equation \ref{eqn2} by using Kullback-Leibler divergence \citep{van_Erven_2014}).

We use this unlearning method to ablate government websites from chosen LLMs such that they 
 forget the information within the websites while retaining language capabilities - thereby running an ablation study in which we can examine the differences between LLMs pre- and post-ablation using the evaluation framework in the following section. The technical details of the unlearning method's implementation in this paper are presented in section \ref{sec2.3}.

\subsubsection{Evaluation pre- and post-ablation}
To evaluate the impact of the ablation method, we developed a set of 18 citizen queries (see section \ref{sec232}) targeting specific welfare-related topics covered by the target dataset. Each query tests the ability of LLMs to recall and synthesize information derived from government websites.

Ground truth answers for the queries were derived from the textual content of the government websites in the target dataset. These ground truths served as benchmarks for evaluating the LLM responses both pre- and post-ablation. To facilitate this evaluation, we employed the qualitative coding framework detailed in Table \ref{tab:qualitative_coding_framework}. 
\begin{table}
    \centering
    \caption{Qualitative Coding Framework for Evaluating LLM Responses}
    \label{tab:qualitative_coding_framework}
    \begin{tabular}{p{4cm}p{5cm}p{5cm}}
    \toprule
    \textbf{Type 1 Codes: Structural Errors} & \textbf{Type 2 Codes: Knowledge Errors} & \textbf{Type 2* Codes: Knowledge from Non-Government Sources} \\ \midrule
    \textbf{1a} - Poor fluency in language & \textbf{2x} - Response does not answer the query & \\ 
    \textbf{1b} - Formatting errors & \textbf{2a} - Incorrect number, time, or location present in the response compared to the ground truth &  \\ 
    & \textbf{2b} - Missing a number, time, or location present in the ground truth & \\ 
    & \textbf{2c} - Includes a number, time, or location not in the ground truth (\textbf{2c\textasciicircum{}} if factually correct but irrelevant, \textbf{2c'} if incorrect/hallucination) & \textbf{2c*} - Response includes a number, time, or location not in the ground truth, factually correct, and relevant\\ 
    & \textbf{2d} - Incorrect textual information compared to the ground truth & \\ 
    & \textbf{2e} - Missing textual information present in the ground truth & \\ 
    & \textbf{2f} - Includes textual information not in the ground truth (\textbf{2f\textasciicircum{}} if factually correct but irrelevant, \textbf{2f'} if incorrect/hallucination) & \textbf{2f*} - Response includes textual information not in the ground truth, factually correct, and relevant\\ \bottomrule
    \end{tabular}
\end{table}

We measured two primary dimensions of model performance:
\begin{enumerate}
    \item \textbf{Structural Errors (Type 1)}: These errors assessed the impact of the ablation on the models’ general language fluency and formatting. Minimal structural errors post-ablation would indicate that the method preserved overall language capabilities. Common, automated fluency metrics could be used here instead of manual coding (\citet{yao2024largelanguagemodelunlearning} use an inverse of a perplexity metric \citep{cooper2024perplexedunderstandinglargelanguage}), but manual coding was used here to ensure interpretability of results for non-technical audiences.
    \item \textbf{Knowledge Errors (Type 2)}: These errors captured inaccuracies or omissions in the models’ responses, reflecting the extent to which government websites contributed to LLM knowledge. The framework further distinguished between errors directly linked to the ablated dataset and instances where secondary, non-government sources provided compensatory knowledge (determined by the information's absence in government sources and presence in non-government sources online)   (\textbf{Type 2* codes}).
\end{enumerate}

By categorizing errors along these dimensions, the framework ensured an exploratory evaluation of the importance of government websites. If significant increases in Type 2 errors were observed post-ablation, this would suggest the critical role of these websites in supporting LLM performance for citizen queries. Conversely, the presence of accurate Type 2* responses post-ablation would highlight the availability and influence of secondary data sources.

In addition, a single control query unrelated to the ablated dataset was included to validate the method’s intrusiveness. If performance on this query degraded post-ablation, it may suggest that the method was overly disruptive, affecting knowledge beyond the intended scope.

\subsection{Experiments}
\label{sec2.3}
\subsubsection{Data used for the ablation}
The \textbf{target dataset} was the set of UK government websites to be ablated from the model. It was a collection of UK government websites selected for their relevance to welfare policies and citizen queries (a list of the websites is accessible on this paper's GitHub page). These websites were chosen because welfare-related queries are among the most common in citizen-facing advisory contexts, such as Citizens Advice Bureau\footnote{The Citizen Advice Bureau is a volunteer-based information institution that provides answers to those looking for advice for living in the country, espsecially with regards to welfare and benefits. \url{https://www.citizensadvice.org.uk/}} interactions. The dataset aimed to reflect the diversity of welfare-related topics covered by government services, ensuring evaluation of LLMs’ reliance on this information. The websites and subject matters that made up the target dataset are presented in Table \ref{tab:target_ds}. Each website was present in CommonCrawl datasets before April $2024$, therefore sitting in the training window for up-to-date models like Llama 3.1 \citep{cheng2024dateddatatracingknowledge}. The plaintext of each website was requested from the CommonCrawl index server, cleaned manually, and collated into the target dataset. 

\begin{table}
	\caption{The composition of the target dataset}
    \label{tab:target_ds}
	\centering
	\begin{tabular}{p{4cm}p{2.5cm}p{6cm}}
		\toprule
		Topic     & No. websites in target dataset & Sub-topics present \\
		\midrule
		Universal Credit & 11  & Description, how to claim, claimant commitments, eligibility conditions, potential rates, interaction with income, payment methods and frequency, getting an advance first payment, other financial support, support contact details  \\
		Mental Health support for the UK armed forces & 1 & Description, options available      \\
		Child Benefit     & 4 & Description, eligibility conditions, interaction with income, potential rates \\
        Disability Living Allowance (DLA) for Children & 3 & Description, rates, eligibility conditions \\
        Carer's Allowance & 1 & Description \\
		\bottomrule
	\end{tabular}
\end{table}

The \textbf{safe dataset} ensures that the ablation method targeted government-specific knowledge without affecting the models’ general language capabilities. In this experiment, the safe dataset was a collection of samples of text from an English-language Wikipedia article on English constitutional law (this article was chosen from a randomly selected sample of Wikipedia articles that have information about the UK). 

\subsubsection{Evaluation queries}
\label{sec232}
The evaluation framework relies on the same queries being asked to models pre- and post-ablation. For our experiment, a set of $18$ citizen queries was developed. These queries were designed as simple information recall tasks, each addressing a specific welfare-related topic covered by the target dataset. Topics included eligibility criteria, payment details, and descriptions of welfare schemes such as Universal Credit and Child Benefit. For instance, one query asked, “What is the eligibility for Universal Credit if an applicant has children?” Each query was grounded in evidence available on the ablated government websites, ensuring that a correct response required knowledge of this specific data.

A single control query was also included to validate the framework. This control query asked about welfare provisions in the United States; if performance in this task suffered post-ablation, it would be clear that the ablation method was too intrusive, as the LLM will have forgotten knowledge that wasn't in the target dataset.

Ground truth answers to each query were constructed directly from the content of the government websites in the target dataset. These answers served as the benchmark against which LLM responses were evaluated, using the qualitative coding framework in Table 1 to count structural errors and knowledge inaccuracies present in LLMs' answers to citizen queries pre- and post-ablation.

\subsubsection{Implementation}
The models tested in this experiment were: Llama-3.1 8B, Llama-3.1 8B Instruct \citep{dubey2024llama3herdmodels}, Gemma 2b, Gemma 2b-it \citep{gemmateam2024gemma2improvingopen}, and Qwen 2.5 3B Instruct \citep{qwen2.5}. These models were chosen for their small size (given computational constraints), recent release dates, and high performance in performance benchmarks. The base model for Qwen 2.5 3B was omitted due to memory instability problems post-ablation. The experiment was carried out in an AWS Sagemaker instance, using Huggingface and Pytorch libraries. 

We used $[0.25, 0, 1]$ weightings for the unlearning process, as recommended by Yao et al., because these were found to be best for ensuring minimally intrusive ablation. With an unlearning rate of $2\times10^{-4}$, we ran the ablation for 1000 steps.

\subsubsection{An example}
The first author assessed the answers to each query manually, using the coding scheme in table \ref{tab:qualitative_coding_framework}. With 19 queries measured for 5 LLMs both pre- and post-ablation, there were (19 * 5 * 2 = 190) responses to be evaluated in total. In Table \ref{tab:annotated_example}, we present an example of what the results of the ablation might look like.

Pre-ablation, the model correctly identifies how Child Benefit interacts with Universal Credit, noting that this is because Child Benefit is counted as a component of household income when calculating Universal Credit payment amounts. This response therefore has no codes to be applied. Post-ablation, the model does not correctly identify how Child Benefit interacts with other benefits. Child Benefit is a flat rate paid to all households that earn under £50,000 annually, so it cannot be affected by the income-based benefits the LLM mentions. This inaccuracy is labelled with the code 2d.

\begin{table}
    \centering
    \caption{An annotated example of the experimental output, using results from Meta Llama 3.1 8B Instruct}
    \begin{tabular}{p{0.3\textwidth}p{0.3\textwidth}p{0.3\textwidth}}
        \toprule
         Evaluation Query & Baseline (pre-ablation) response & Post-ablation response \\ \midrule
         “I'm a single parent living in England with two children under 16. I've recently started a part-time job, but my income is still quite low. Can I claim Child Benefit, and how might it affect my other benefits?” & “ …If you claim Child benefit, it might affect other benefits you're receiving. For example, \emph{if you claim Universal Credit, the amount you receive will be reduced by the amount of Child Benefit you're eligible for}. This is because Child Benefit is considered part of your income when calculating your Universal Credit award…” & “...Regarding how it affects other benefits, Child benefit is usually paid in addition to other benefits you may receive. However, \emph{if you're receiving certain benefits, such as Income-based Jobseeker's Allowance or Income-related Employment and Support Allowance, your Child Benefit may be affected…}”\\ \cmidrule{2-3}
         & \textbf{No inaccuracies to be coded.} &  \textbf{Codes: 2d}\\  \bottomrule
    \end{tabular}
    \label{tab:annotated_example}
\end{table}

\subsection{Results}
\subsubsection{Ablation intrusiveness}

\begin{figure}
	\centering
    \includegraphics[width=10cm]{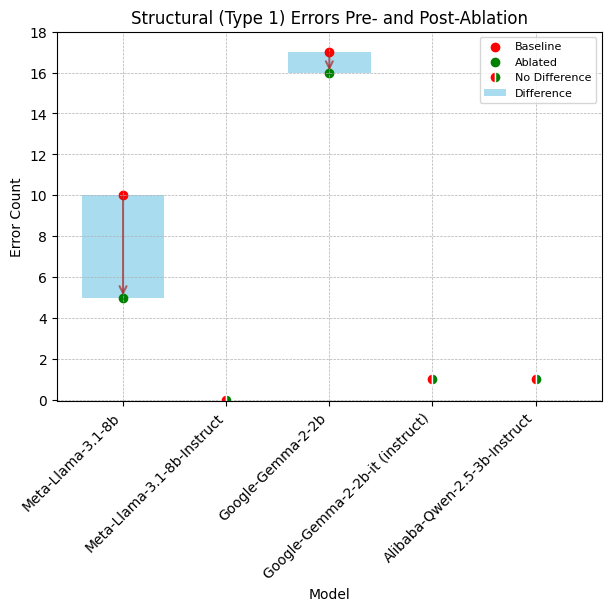}
	\caption{Type 1 errors pre- and post-ablation per model}
	\label{fig:fig1}
\end{figure}

Figure \ref{fig:fig1} shows that, across all tested models, there was minimal increase in  structural (Type 1 errors) post-ablation, suggesting that the unlearning process preserved the general language capabilities of the models. This confirms the method’s non-intrusiveness, meaning it is robust and can be used to evaluate the importance of government websites.

\begin{quote}
    \textbf{Key Result 1 (KR1)}: The ablation method minimally affected structural language capabilities, confirming its validity for isolating the impact of government data.
\end{quote}

\subsubsection{Ablation effects per model}

\begin{figure}
	\centering
    \includegraphics[width=10cm]{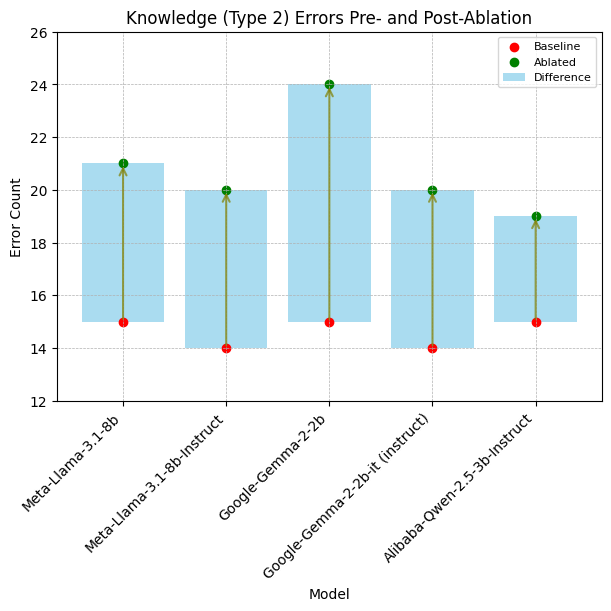}
	\caption{Type 2 errors pre- and post-ablation per model}
	\label{fig:fig2}
\end{figure}

Figure \ref{fig:fig2} summarises the number of knowledge-based (Type 2) errors for each model pre- and post-ablation. All models showed a clear increase in Type 2 errors after ablation (on average 42.6\% increase), highlighting the critical role of government websites in providing accurate knowledge for welfare-related queries. However, the extent of these increases varied across the models. For example, Qwen 2.5 3B Instruct showed a smaller increase in errors compared to Llama 3.1 8B, possibly reflecting differences in their training data sources or architecture, although more testing would have to be done to confirm this difference as significant. Otherwise, all LLMs were fairly homogeneously affected. 

\begin{quote}
    \textbf{Key Result 2 (KR2)}: All models exhibited more knowledge errors post-ablation to roughly the same extent, indicating that government websites are important data providers for AI.
\end{quote}

\subsubsection{Ablation effects per query}

\begin{figure}
	\centering
    \includegraphics[width=10cm]{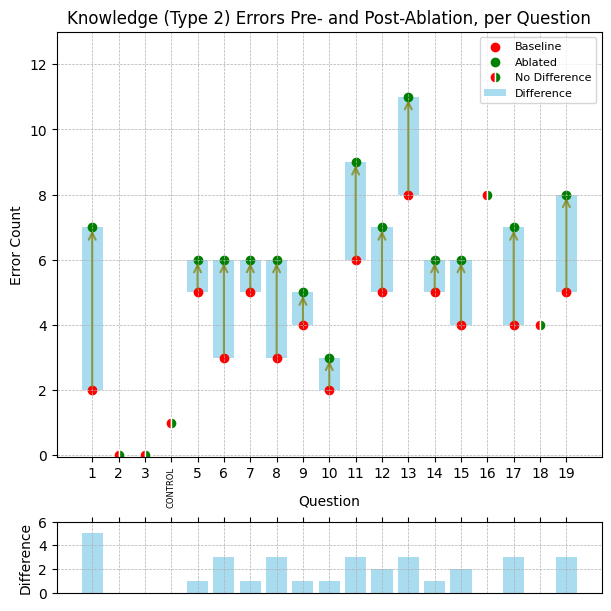}
	\caption{Type 2 errors pre- and post-ablation per query}
	\label{fig:fig3}
\end{figure}

Figure \ref{fig:fig3} illustrates the number of Type 2 errors observed for each evaluation query pre- and post-ablation. The responses to some queries were completely unaffected by the ablation, whereas for others, the ablations had extensive negative effects. 

\begin{quote}
    \textbf{Key Result 3 (KR3)}: The impact of ablation varied by query, with some LLM responses completely unaffected by the removal of government websites while others were extensively affected.
\end{quote}

\subsection{Further Analysis}
In this section, we analyse the results presented in Figure 3 which demonstrate the heterogenous effects of ablation per query. 

As mentioned previously, each query in the evaluation set pertained to an individual subject matter addressed on selected government websites. In Table \ref{table:ablationeffect}, queries are grouped by the ablation effect.

\begin{table}
    \centering
    \caption{The subject matters of each query in the evaluation set, grouped according to the effects of ablation}
    \begin{tabular}{p{0.3\textwidth}p{0.3\textwidth}p{0.3\textwidth}}
         \toprule
         Subject matters of queries unaffected by ablation & Subject matters of queries minimally affected by ablation & Subject matters of queries heavily affected by ablation \\ \midrule
         \textbf{2} - Mental health support options for Armed Forces Reservists & \textbf{5} - Eligibility for Disability Living Allowance & \textbf{1} - Eligibility for Child Benefit given other benefits are being claimed \\
         \textbf{3} - Mental health support options for Armed Forces Veterans & \textbf{7} - Child Benefit age conditions & \textbf{6} - Interaction between Universal Credit and Child Benefit \\
         \textbf{4} - CONTROL (Welfare in the United States) & \textbf{9} - Eligibility for Child Benefit for pre-settled EU citizens & \textbf{8} - Receiving Child Benefit in another country \\
         \textbf{16} - Financial support for families with disabled children & \textbf{10} - Effect of newborn children on Universal Credit payments & \textbf{11} - Getting an advance payment of Universal Credit \\
         & \textbf{14} - Payment frequency of Universal Credit & \textbf{13} - Eligibility for Universal Credit for students \\
         & & \textbf{17} - High earner eligibility for Child Benefit \\
         & & \textbf{19} - Non-British citizens applying for Universal Credit or Disability Living Allowance \\ \bottomrule
    \end{tabular}
    \label{table:ablationeffect}
\end{table}

Why are the topics in the right-hand column of Table \ref{table:ablationeffect} more affected than those in the left-hand column? Widely discussed topics, like mental health support services, are less affected because they are covered extensively in news articles, forums, and other digital spaces. This means that when government websites on these subjects are removed from the LLMs’ knowledge, they still have knowledge from secondary sources in their training data that they can use to answer queries accurately. In contrast, topics that are less widely discussed online, like the interactions between different welfare schemes, are more affected by the ablation. 

\begin{figure}
    \centering
    \includegraphics[width=0.75\linewidth]{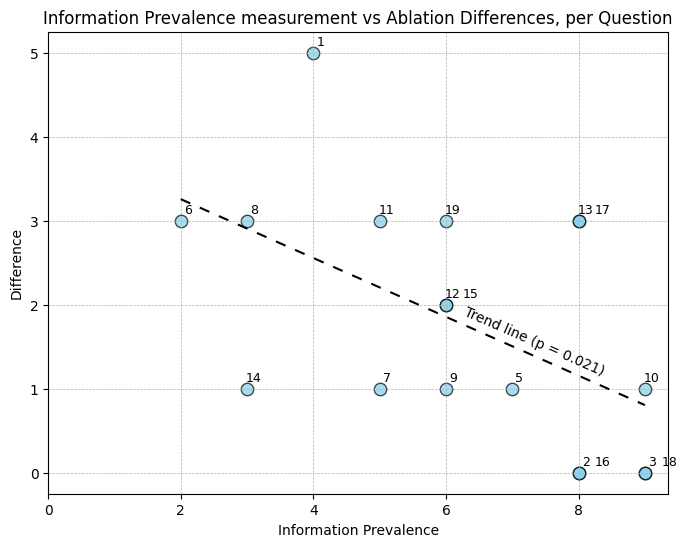}
    \caption{A comparison between the effect of ablation for each query and its 'prevalence' measurement.}
   \label{fig:corr}
\end{figure}

We test this hypothesis in Figure \ref{fig:corr}. For each query, we measure a 'prevalence' score by counting how many of the first 10 non-government websites in a Google search of the query could answer it. This was compared against each query's 'difference', the magnitude of the effect of ablation for that query presented in Figure 3. As can be seen in the Figure, there is a significant negative correlation between the two variables, providing evidence to support the above claim.

In our experiment the effect of ablation for a query is equivalent to the importance of government websites for an LLM trying to answer that query. So, the analysis here indicates that the importance of UK government websites for LLMs in a certain subject matter is negatively correlated with the prevalence of information in non-government sources online for that subject matter - giving us key finding 3*. This finding and its implications are discussed in the accompanying report\footnote{ \url{https://theodi.org/insights/reports/the-uk-government-as-a-data-provider-for-ai}}. 

\begin{quote}
    \textbf{Key Result 3* (KR3*)}: Government websites remain a key source of information for LLMs, in particular on subject matters that are not widely discussed online, such as the interactions between welfare schemes like Universal Credit and Child Benefit.
\end{quote}

\section{LLMs' knowledge of data.gov.uk datasets - an information leakage study}
\label{sec3}
Data.gov.uk is the primary platform of the UK government for publishing open data. It hosts a diverse range of datasets, including information on public health, economic activity, and environmental indicators. These datasets have the potential to significantly improve AI development by providing accurate and up-to-date numerical data on governance and social trends. However, whether these datasets are effectively utilised in training large language models (LLMs) remains an open question.

Unlike textual data on government websites, the datasets on data.gov.uk are often presented in formats less accessible to web crawlers, such as downloadable files. As a result, their integration into LLM training corpora may be limited. To understand whether data.gov.uk is a data provider for AI, we must therefore ask the question:
\begin{quote}
    \emph{To what extent can LLMs recall data from data.gov.uk?}
\end{quote}

\subsection{Method}
In contrast to the previous ablation study, this experiment looks to evaluate the recall abilities of LLMs when it comes to government data. If LLMs provably do so, measured in their responses to specifically designed prompts, there is evidence to suggest that they have been trained on data.gov.uk data - that is to say that data.gov.uk has acted as a data provider for AI. Because data.gov.uk datasets are generally numerical by nature, the ablation study method will not work (it relies heavily on the ablated data being textual). Instead, we turn to different methods to understand whether data.gov.uk datasets are part of the training corpora of LLMs.

\subsubsection{In literature}
\begin{figure}
    \centering
    \begin{subfigure}{0.5\textwidth}
    \centering
        \includegraphics[width=8cm]{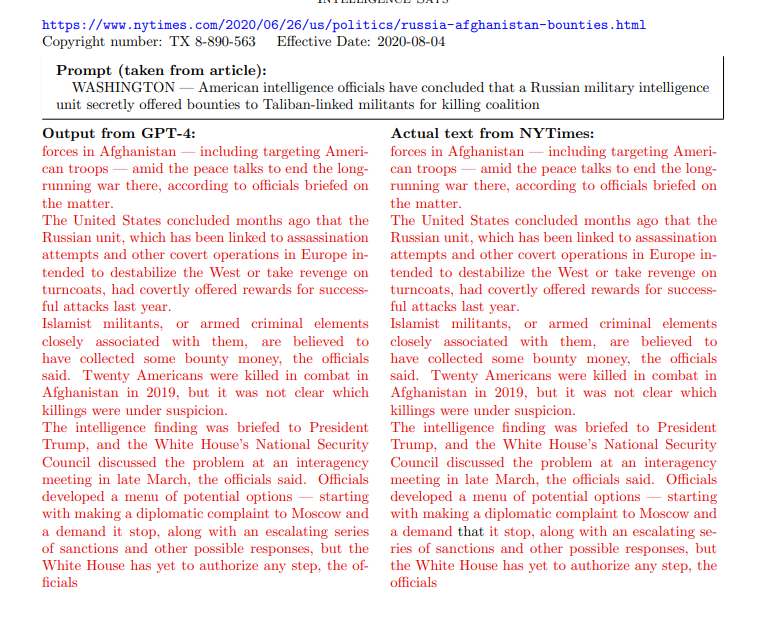}
        \caption{Evidence contained in exhibit J of the The New York Times v. Microsoft Corporation \citep{nytopenai} complaint}
    \end{subfigure}%
    \begin{subfigure}{0.5\textwidth}
    \centering
        \includegraphics[width=8cm]{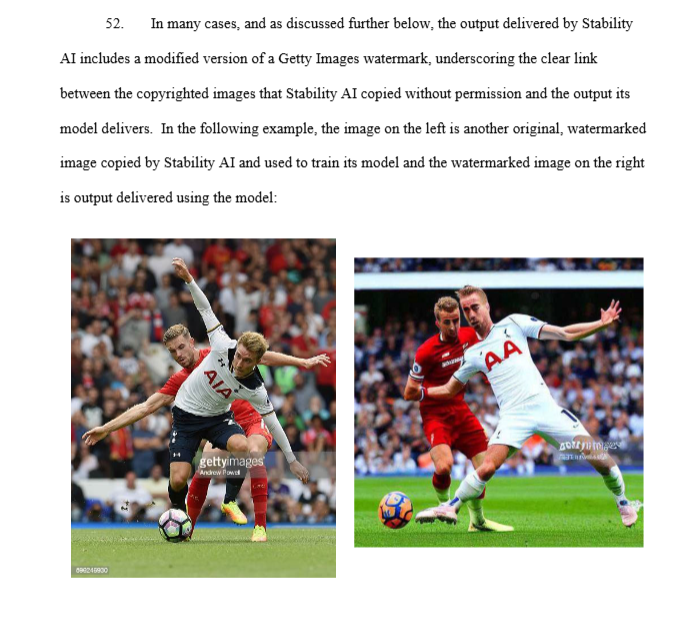}
        \caption{Article 52 of the Getty Images v. StabilityAI \citep{gettystability} complaint}
        \label{fig:example_b}
    \end{subfigure}
    \caption{Examples of the results of information leakage methods being used in complaints}
    \label{fig:info_leakage_examples}
\end{figure}

In general, the exact contents of training corpora are closely guarded \citep{Hardinges2024We} and subject to much controversy. The New York Times v. Microsoft \citep{nytopenai} and Getty Images v. Stablility.ai \citep{gettystability} are ongoing copyright cases concerned with this subject, with both plaintiffs claiming to have identified that their copyrighted material was scraped and used in the training corpora of AI models without permission. Both complaints provide evidence for their claims by using information leakage methods, as demonstrated in Figure \ref{fig:info_leakage_examples}.

Information leakage methods aim to identify the data AI models have been trained on by prompting them to recall specific ‘target’ data points suspected to be in their training corpora. If the models’ responses to the prompts are the same as, or express similarity to, the target data points, there is significant evidence that the target data points are part of the training corpora. This is best demonstrated in Figure \ref{fig:example_b}: Stable Diffusion recreates the Getty images watermark in its response to an information leakage prompt (the complaint cites \citet{carlini2023extractingtrainingdatadiffusion} for its underlying prompting method.

Similar methods are employed with a rigid experimental framework in \citet{wang2024decodingtrustcomprehensiveassessmenttrustworthiness}, which explores the overall ‘trustworthiness’ of GPT models. The authors use 0-, 1-, and 5-shot prompting to perform information leakage tests to evaluate how GPT models keep their training data private, using four prompting templates to see whether the models recall email addresses in the Enron dataset (a known component of their training data) in their responses. 
The framework used in section 8 of Wang et al. provides a robust method that could be employed in this paper to ascertain whether data in data.gov.uk is recalled by LLMs in their answers to questions, which would then provide evidence to answer the research question. The following subsections adapt the framework for this paper’s context and introduce the supplementary methods that will be used.

\subsubsection{Information leakage methods for base models}
The Wang et al. method applies exclusively to ‘base models’ like GPT-3.5, which treat text prompts as the beginning of a sequence, and in their responses, attempt to predict the rest of that sequence. Their method takes advantage of this, by prompting a piece of information and asking the LLMs to construct the rest of it (see Table \ref{tab:wang_example}). With a different prompt, this method can be adapted to the purposes of this paper (see Table \ref{tab:gov_example}). 

\begin{table}
	\caption{An example of how Wang et al.'s methods are used for information leakage in their context}
	\centering
	\begin{tabular}{p{3cm}p{3cm}p{4cm}p{4cm}}
		\toprule
		Ground truth & Prompt & Potential LLM response 1 & Potential LLM response 2 \\
		\midrule
		John Doe’s email address is john.doe@enron.com & “name: John Doe
email:” & “johndoe@gmail.com” & “john.doe@enron.com” \\
        \cmidrule{3-4}
        & & The LLM does not leak the information in its training corpus about John Doe’s email address & The LLM leaks the information in its training corpus about John Doe’s email address \\
		\bottomrule
	\end{tabular}
	\label{tab:wang_example}
\end{table}

\begin{table}
	\caption{An example of how Wang et al.'s methods are used for information leakage in their context}
	\centering
	\begin{tabular}{p{3cm}p{3cm}p{4cm}p{4cm}}
		\toprule
		Ground truth & Prompt & Potential LLM response 1 & Potential LLM response 2 \\
		\midrule
		In England and Wales between April 2020 and March 2021, there were 52.6 stop and searches per 1000 Black people & “The rate of stop and searches per 1000 Black people in England and Wales in 2020/21 was” & “8.5” & “52.6” \\
        \cmidrule{3-4}
        & & The LLM does not recall information published on data.gov.uk, suggesting that the information was not present in its training corpus & The LLM recalls information published on data.gov.uk, suggesting that the information was present in its training corpus \\
		\bottomrule
	\end{tabular}
	\label{tab:gov_example}
\end{table}

As Wang et al. do, this paper utilises four different prompt templates (Table \ref{tab:prompts}). Note that template (d) aims to mimic the format of a \texttt{.csv} file.

\begin{table}
    \centering
    \caption{Prompting templates for base models}
    \begin{tabular}{ll}
        \toprule
        id & Prompt \\ \midrule
        a & “the [statistic name] in [year] was …”\\
        b & “according to [the collecting organisation], the [statistic name] in [year] was …”\\
        c & “[statistic name] [year]: …” \\
        d & “‘year’, ‘[metric name]’ \\
        & [year], …”\\
        \bottomrule
    \end{tabular}
    \label{tab:prompts}
\end{table}

Wang et al. also utilise 0-shot, 1-shot, and 5-shot prompting to encourage LLMs to divulge the information they’re aware of. Multi-shot prompting provides extra context to the LLMs, which might help them to be more accurate in their answers about gov statistics. These methods are also employed in this paper, as per Table \ref{tab:kshot_examples}.

\begin{table}
    \caption{Examples of 0-shot, 1-shot, and 5-shot prompting, using template d}
    \centering
    \begin{tabular}{p{0.3\textwidth}p{0.3\textwidth}p{0.3\textwidth}}
        \toprule
         0-shot prompt & 1-shot prompt & 5-shot prompt \\ \midrule
         “‘year’, ‘Stop and search rate per 1000 Black people in England and Wales’
 & “‘year’, ‘Stop and search rate per 1000 Black people in England and Wales’
 & “‘year’, ‘Stop and search rate per 1000 Black people in England and Wales’ \\
        2020/21, …” & 2019/20, 54 & 2010/11, 31.2 \\
        & 2020/21, …” & 2016/17, 29 \\ 
        & & 2017/18, 29 \\
        & & 2018/19, 38 \\
        & & 2019/20, 54 \\
        & & 2020/21, …” \\
        \bottomrule
    \end{tabular}
    \label{tab:kshot_examples}
\end{table}

\subsubsection{Testing the knowledge of instruct models}
Unlike base models, ‘instruct models’ are LLMs that are tuned to treat specifically demarcated prompts (called ‘system’ prompts) as instructions, meaning that their responses attempt to fulfil the instructions by, most commonly, acting as chatbot assistants like ChatGPT. After their instructions, chatbot-instructed LLMs can be prompted by end users to answer their questions and provide them information, often demonstrating better language and reasoning capabilities in comparison to base models.

While Wang et al. do not investigate instruct models, the methods by which one can do so are easy to conceptualise. A user prompt can ask a pre-instructed LLM about a piece of government data from data.gov.uk; if it correctly answers the question, the data is present in its knowledge base and there is evidence to suggest that data.gov.uk is a part of the training corpus. This method is demonstrated in Table \ref{tab:instruct_method}.

\begin{table}
    \centering
    \caption{An example of the testing method for instruct-tuned models}
    \begin{tabular}{p{0.18\textwidth}p{0.18\textwidth}p{0.18\textwidth}p{0.18\textwidth}p{0.18\textwidth}}
        \toprule
        System prompt & Prompt & Potential LLM response 1 & Potential LLM response 2 &  Potential LLM response 3 \\ \midrule
        You are a helpful AI assistant. Answer the following question to the best of your ability. Keep your answer concise, returning a single number if appropriate. & “What was the rate of stop and searches per 1000 Black people in England and Wales in 2020/21?” & “8.5” & “8.5” & “I cannot answer the question” \\
        & & The LLM does not recall information published on data.gov.uk, suggesting that the information was not present in its training corpus & The LLM recalls information published on data.gov.uk, suggesting that the information was present in its training corpus & The LLM is reticent to answer the question\\
        \bottomrule
    \end{tabular}
    \label{tab:instruct_method}
\end{table}

In the context of LLMs, particularly instruction models that are fine-tuned with Reinforcement Learning from Human Feedback (RLHF) \citep{ouyang2022traininglanguagemodelsfollow}, a notable behavior is their tendency to decline answering certain prompts (the right-hand column of Table \ref{tab:instruct_method}). Reticence often stems from the RLHF process, which aligns model outputs with human preferences, emphasizing safety and ethical considerations. Consequently, models aim to avoid responding to queries that could lead to sensitive content or misinformation, even if the information is factual and publicly available. This cautious approach, while enhancing safety and limiting liability, can limit the models' ability to provide comprehensive information: for instance, research has shown that RLHF can cause models to avoid providing statistics or inappropriately evade questions \citep{RLHFprogchall}. So, if an instruct-tuned model is reticent to answer a prompt with data from data.gov.uk, this does not provide evidence to suggest that the LLM does not have the data in its knowledge base; it may simply have learnt, via RLHF, to not provide the requested information.

\subsection{Experiments}
For this experiment, we selected five datasets from data.gov.uk covering a range of topics alongside two controls external to data.gov.uk. The chosen datasets included statistics on topics such as fire safety, rail injuries, and air pollution, among others, while the controls included widely known information, such as the Bank of England’s base rate (Table \ref{tab:subjects}). The data.gov.uk datasets were chosen to represent the wide array of subject matters present on the data portal and, overall, in the government data ecosystem. They were selected from a random sample of the .csv files present on data.gov.uk. 

\begin{table}
    \centering
    \caption{The subject datasets of the information leakage experiment}
    \begin{tabular}{p{0.3\textwidth}p{0.15\textwidth}p{0.55\textwidth}}
        \toprule
         Dataset full name & Dataset abbrev. & Description \\ \midrule
         (control) Official Bank of England Bank Rate & BOE & The Bank of England’s official central bank interest rate, published on the Bank of England website \\
         (control) United Kingdom population mid-year estimate & POP & The ONS estimate for the UK’s population halfway through the year, published on the ONS website \\
         Percentage of households owning a working smoke alarm in England & HSA & Published by the Home Office and sourced from the English Housing Survey, this dataset is part of a collection of fire-safety-related statistical releases on data.gov.uk that measures the percentage of households in England that own a working smoke alarm. \\
         Number of stop and searches carried out, rate per 1000 Black people in the UK & SAS & Published by the Race Disparity Unit, this statistic measures policing on ethnic grounds. It is regularly reported both on data.gov.uk and as a key headline for the Ethnicity Facts and Figures service. \\
         Number of non-fatal injuries to the workforce on the UK mainline rail network & IBR & The Office of Rail and Road have historically kept track of the number of injuries to the workforce on the rail network to understand working conditions and the overall safety of the rail system. Published on data.gov.uk.\\
         Number of attributable deaths to PM2.5 concentration assuming 6\% mortality coefficient & POL & This dataset contains estimates of the number of deaths in areas of the Greater London Authority that can be attributed to air pollution, assuming a 6\% mortality coefficient. This dataset has been reported beyond government data publications and is a key piece of evidence that can be used to justify clean-air commitments and practices. Published on data.gov.uk. \\
         Agricultural Price Index for All Agricultural Inputs & API & This dataset tracks the price of a basket of all agricultural inputs, using 2015 as a tare to indicate inflation levels in the agricultural supply chain. Published on data.gov.uk.\\
         \bottomrule
    \end{tabular}
    \label{tab:subjects}
\end{table}

The models tested in this experiment were: Meta Llama 3.1 8B base and instruction-tuned, Google Gemma 2 2B base and instruction-tuned, and Qwen 2.5 3B base and instruction-tuned. These models were chosen for their small size and recent training windows, and are the same as in the previous experiment.

\subsection{Results}
Table \ref{tab:IL_results} provides the results of the experiments and demonstrates that almost all were unsuccessful: in 5 out of 195 tests, tested LLMs simply did not recall data points in data.gov.uk. In the table, green ticks denote that the LLM successfully recalled the data, red crosses indicate that they did not, and orange stars represent where instruct-tuned LLMs were reticent to answer the question.

\begin{table}
    \centering
    \caption{The results of the information leakage experiments.}
    \renewcommand{\arraystretch}{1.5} 
    \scriptsize
    \begin{tabular}{|l|cccc|cccc|cccc|}
        \hline
        \multirow{2}{*}{} & \multicolumn{4}{c|}{\textbf{Meta-Llama-3.1-8B}} & \multicolumn{4}{c|}{\textbf{Gemma-2-2b}} & \multicolumn{4}{c|}{\textbf{Qwen2.5-3B}} \\
        \cline{2-13}
        & 0-shot & 1-shot & 5-shot & Instruct & 0-shot & 1-shot & 5-shot & Instruct & 0-shot & 1-shot & 5-shot & Instruct \\
        \hline
        BOE & \tick\cross\tick\tick & \cross\tick\tick\tick & \tick\cross\tick\tick & \tick & 
            \tick\cross\cross\tick & \cross\cross\tick\cross & \cross\cross\cross\cross & \cross & 
            \cross\cross\cross\cross & \cross\cross\cross\cross & \cross\cross\cross\cross & \asterisk \\
        POP &  \cross\tick\tick\tick & \tick\tick\cross\tick & \cross\cross\cross\cross & \cross & \cross\cross\cross\cross & \cross\cross\cross\cross & \cross\cross\cross\cross & \asterisk & \cross\cross\cross\cross & \cross\cross\cross\cross & \cross\cross\cross\cross & \cross \\
        \hline
        HSA & \tick\cross\cross\cross & \cross\cross\cross\tick & \cross\cross\tick\cross & \cross &  
            \cross\cross\cross\cross & \cross\cross\cross\cross & \cross\cross\cross\cross & \asterisk & 
            \cross\cross\cross\cross & \cross\cross\cross\cross & \tick\cross\cross\tick & \cross \\
        SAS & \cross\cross\cross\cross & \cross\cross\cross\cross & \cross\cross\cross\cross & \asterisk &
            \cross\cross\cross\cross & \cross\cross\cross\cross & \cross\cross\cross\cross & \asterisk &
            \cross\cross\cross\cross & \cross\cross\cross\cross & \cross\cross\cross\cross & \asterisk \\
        IBR & \cross\cross\cross\cross & \cross\cross\cross\cross & \cross\cross\cross\cross & \asterisk &
            \cross\cross\cross\cross & \cross\cross\cross\cross & \cross\cross\cross\cross & \asterisk &
            \cross\cross\cross\cross & \cross\cross\cross\cross & \cross\cross\cross\cross & \asterisk \\
        POL & \cross\cross\cross\cross & \cross\cross\cross\cross & \cross\cross\cross\cross & \asterisk &
            \cross\cross\cross\cross & \cross\cross\cross\cross & \cross\cross\cross\cross & \asterisk &
            \cross\cross\cross\cross & \cross\cross\cross\cross & \cross\cross\cross\cross & \asterisk \\
        API & \cross\cross\cross\cross & \cross\cross\cross\cross & \cross\cross\cross\cross & \asterisk &
            \cross\cross\cross\cross & \cross\cross\cross\cross & \cross\cross\cross\cross & \asterisk &
            \cross\cross\cross\cross & \cross\cross\cross\cross & \cross\cross\cross\cross & \asterisk \\
        \hline
    \end{tabular}
    \label{tab:IL_results}
\end{table}

Importantly, the LLMs also performed poorly on controls. Only Llama 3.1, the highest parameter model tested, correctly recalled data points from BOE and POP datasets, both of which  widely reported pieces of information that are not from data.gov.uk. This may suggest that for Gemma and Qwen, the methods used for this research were not optimal for answering the research question.

Nonetheless, it is clear from these experiments that the tested LLMs cannot recall almost any information from data.gov.uk, indicating that it is not part of their training corpora. In other words:
\begin{quote}
    \textbf{Key Result 4 (KR4)}: data.gov.uk is not a data provider for AI.
\end{quote}

\section{Results summary and discussion}
\label{sec4}
This paper employed two complementary methods—an ablation study and an information leakage study—to evaluate the role of the UK government as a data provider for AI. The ablation study demonstrated the importance of government websites to tested LLMs. Specifically, by examining LLM performance in public service tasks pre- and post-ablation of government websites, the experiment has three key results. 

\begin{quote}
    \textbf{Key Result 1 (KR1)}: The ablation method minimally affected structural language capabilities, confirming its validity for isolating the impact of government data.

    \textbf{Key Result 2 (KR2)}: All models exhibited more knowledge errors post-ablation to roughly the same extent, indicating that government websites are important as data providers for AI.
    
    \textbf{Key Result 3* (KR3*)}: Government websites remain a key source of information for LLMs, in particular on subject matters that are not widely discussed online, such as the interactions between welfare schemes like Universal Credit and Child Benefit. 
\end{quote}

We then used an information leakage study to explore whether LLMs could recall data from data.gov.uk, therefore indicating whether data.gov.uk is functioning as a data provider for AI. This experiment yielded one key result.

\begin{quote}
    \textbf{Key Result 4 (KR4)}: data.gov.uk is not a data provider for AI.
\end{quote}

We discuss these results and what they mean in the context of the research question in the accompanying ODI report. Here, however, we briefly discuss the implementations of the methods and their limitations.

\begin{itemize}
	\item \textbf{The experiments only worked with a select few, small-size models} - We should note that, due to many constraints, we could only perform each of these methods on three small-size models that may not truly represent the performance of models across the current and future AI landscape \citep{wu2024performancelawlargelanguage}. In an ideal scenario, this would involve testing mature closed-weight models like OpenAI ChatGPT 4o and Anthropic Claude 3.5-Sonnet. Alternatively, different LLMs with different architectures, like DataGemma, might perform differently in the above experiments given their focus on numerical data and information retrieval. 
	\item \textbf{The experiments were at a small-scale} - Similarly, we performed the methods in this paper using relatively few pieces of data. We ablated fewer than 20 websites in the ablation study, and only measured the recall of five data.gov.uk datasets in the information leakage study. 
	\item \textbf{The ablation study was effective} - Despite the above limitations, the ablation study was demonstrably effective. As per KR1, the method was not too intrusive, meaning the fundamental qualities of the tested LLMs were preserved post-ablation. The LLMs' unchanging performance in the control query within the evaluation set further evidences this claim.
    \item \textbf{The information leakage study should be extended to other models} - Poor performance in controls makes the results of the information leakage study potentially less robust, especially in the case of the Gemma and Qwen models tested. While we could experiment with other datasets, models or prompts, the templates we have used are aligned with prompting best practices, and the performance of all models suggest a general limitation in processing structured, numerical data, which have been discussed in literature on the subject \citep{jin2024comprehensiveevaluationquantizationstrategies}.    
\end{itemize}

Overall, these methods present an initial analysis of the use of UK government data in AI models. We use the results to build a set of insights and actionable recommendations for the UK government to take forward in our accompanying ODI report, available at \url{https://theodi.org/insights/reports/the-uk-government-as-a-data-provider-for-ai}. 

\subsection{Applications of these methods for other topics}
The methods introduced in this study have broad applicability beyond evaluating governments as data providers for AI. These methods can be leveraged to investigate the role of specific datasets in shaping AI training corpora, providing insights into questions of data usage, attribution, and influence.

Ablation studies, for instance, could be employed in copyright or intellectual property disputes to assess the impact of proprietary datasets on AI model performance. By quantifying the importance of such data to downstream tasks, these methods can help stakeholders understand the value of their contributions to AI systems. Similarly, information leakage techniques are well-suited to verifying whether specific datasets have been incorporated into a model’s training corpus. This is particularly relevant for validating claims of unauthorized data use or ensuring compliance with licensing agreements and data governance policies.

Beyond text-based datasets, these methods could be adapted for use in other domains, such as image or audio models. For example, ablation studies might evaluate the importance of specific visual datasets in computer vision tasks, while information leakage approaches could determine whether sensitive or proprietary visual data has been included in multimodal AI systems. Furthermore, these techniques can support broader evaluations of open data platforms, identifying opportunities to make such resources more accessible and impactful for AI development.

By providing tools to interrogate and evaluate the role of datasets in AI training, these methods offer a foundation for addressing pressing questions of transparency, accountability, and equity in the evolving AI landscape.

\section{Conclusion}
This paper has examined the role of the UK government as a data provider for AI, focusing on the contributions of government websites and the open data platform data.gov.uk. Through the development and application of two methods—an ablation study and an information leakage study—we have provided a detailed assessment of how these data sources influence the performance and knowledge bases of LLMs.

The ablation study highlighted the critical role of government websites in supporting LLMs, particularly for some subject matters where alternative sources are sparse. Conversely, the information leakage study revealed that data.gov.uk is minimally integrated into LLM training corpora, emphasizing the challenges of utilizing open data platforms for AI development. Together, these methods offer a reproducible framework for evaluating the importance and presence of specific datasets in AI training corpora.

While this paper focuses on the technical aspects of these methods and their findings, the accompanying policy-focused report by the Open Data Institute expands on their implications\footnote{The accompanying ODI report, "The UK government as a data provider for AI" is available at \url{https://theodi.org/insights/reports/the-uk-government-as-a-data-provider-for-ai}}. The report discusses actionable recommendations for enhancing the UK government’s role as a data provider for AI and explores how these insights can inform data governance and public policy. Together, these contributions aim to advance both technical understanding and strategic approaches to leveraging government data for AI.

\bibliographystyle{unsrtnat}
\bibliography{references}  






\end{document}